\documentstyle[11pt,moriond,epsfig]{article}

\def\Journal#1#2#3#4{{#1} {\bf #2}, #3 (#4)}

\def\NIMB{{\em Nucl. Instrum. Methods} B}

\def\PLB{{\em Phys. Lett.}  B}
\def\PRL{\em Phys. Rev. Lett.}
\def\PRD{{\em Phys. Rev.} D}

\def\be{\begin{equation}}
\def\ee{\end{equation}}
\def\bea{\begin{eqnarray}}
\def\eea{\end{eqnarray}}

\begin{document}
\vspace*{4cm}
\title{NA48: Rare Decay Results}

\newcommand{\beq}  {\begin{equation}}
\newcommand{\eeq}  {\end{equation}}
\newcommand{\ks}{K_{S}}
\newcommand{\kl}{K_{L}}
\newcommand{\kgg}{K~\rightarrow~\gamma\gamma}
\newcommand{\ksgg}{K_{S}~\rightarrow~\gamma\gamma}
\newcommand{\kspp}{K_{S}~\rightarrow~\pi^{0}\pi^{0}}
\newcommand{\klgg}{K_{L}~\rightarrow~\gamma\gamma}
\newcommand{\klppg}{K_{L}~\rightarrow~\pi^{+}\pi^{-}\gamma}
\newcommand{\kslppee}{K_{S,L}~\rightarrow~\pi^{+}\pi^{-}e^{+}e^{-}}
\newcommand{\ksppee}{K_{S}~\rightarrow~\pi^{+}\pi^{-}e^{+}e^{-}}
\newcommand{\klppee}{K_{L}~\rightarrow~\pi^{+}\pi^{-}e^{+}e^{-}}
\newcommand{\klpppd}{K_{L}~\rightarrow~\pi^{+}\pi^{-}\pi^{0}_{D}}
\newcommand{\kspee}{K_{S}~\rightarrow~\pi^{0}e^{+}e^{-}}
\newcommand{\kspdpd}{K_{S}~\rightarrow~\pi^{0}_{D}\pi^{0}_{D}}
\newcommand{\ksppd}{K_{S}~\rightarrow~\pi^{0}\pi^{0}_{D}}
\newcommand{\klpee}{K_{L}~\rightarrow~\pi^{0}e^{+}e^{-}}
\newcommand{\lnp}{\Lambda~\rightarrow~n\pi^{0}}

\author{ T. \c{C}uhadar-D\"{o}nszelmann\\
On behalf of the NA48 Collaboration\footnote{
Cagliari-Cambridge-CERN-Dubna-Edinburgh-Ferrara-Firenze-Mainz-Orsay-Perugia-Pisa-Saclay-Siegen-Torino-Vienna-Warsaw} }

\address{CERN, Div. EP, CH-1211, \\
Geneva 23, Switzerland}

\maketitle\abstracts{ Recent results on the kaon rare decays of the 
$\ksgg$, $\kslppee$, $\kspee$ measured in NA48 experiment at CERN
are presented in this paper.
}

\section{The experiment}

The NA48 experiment is designed to measure direct CP violation in
two-pion decays and to search for rare decays of neutral kaons
using simultaneous and almost collinear $\ks$ and $\kl$ beams. 
The $\kl$ beam is produced
by 450 GeV/c protons from SPS  hitting a beryllium target an
angle of 2.4 mrad.
The proton intensity is $1.5\times10^{12}$ per SPS pulse and
these pulses are
2.4 s long with a repetition period of 14.4 s. Non-interacting protons 
are directed to the $\ks$ target, which is positioned $\sim120$ m downstream
of the $\kl$ beam line, to produce the $\ks$ beam.
Anticounter at the
end of the final collimator(AKS) defines the beginning of the fidual volume of
$\sim40$ m long. A set of seven scintillator counters(AKL) surround
the decay region and vetoes events outside the detector acceptance.
The energy and the
position of the electromagnetic
showers of electrons and photons as the products of 
the kaon decays are detected in the liquid krypton calorimeter(LKR), with an 
energy resolution of 
$\sigma(E)/E \simeq 0.100/E \oplus 0.032/\sqrt{E} \oplus 0.005$
(E in GeV).
The momenta of charged
particles are measured in a spectrometer consisting of a dipole
magnet and four drift chambers(DCH). The spectrometer has a momentum resolution of 
$\sigma(p)/p \simeq  [0.009{p} \oplus 0.5]\%$ (p in GeV).
A hodoscope placed in front of the calorimeter is used to measure the event
time with resolution of 150 ps. An iron-scintillator hadron
 calorimeter(HAC) is used to measure the energies of the hadrons 
and the muons are identified by an muon veto system. The detailed
description of the experiment can be found elsewhere~\cite{na48}. 

The results presented here are based on data taken in 1998 and 1999 
During
two days in 1999, data was recorded with a 
factor $\sim200$ higher beam intensity than the usual $K_{S}$ beam, resulting
in $2.3\times10^8$ $K_{S}$ decays.

\section{Decays of $\ksgg$}

In the framework of the chiral perturbation theory($\chi$PT),
the contributions of one loop of charged pions and kaons
to $\ksgg$ decays are finite and give an unambigious prediction
for branching ratio, $2.25\times10^{-6}$ with less than $10\%$ uncertainty~\cite{ggth}.
Thus, a precise determination of branching ratio provides a
good test of $\chi$PT. The CERN experiment NA31~\cite{ggex}
measured a value compatible with theory:
$(2.4\pm0.9)\times10^{-6}$.

\begin{figure}[ht]
\psfig{figure=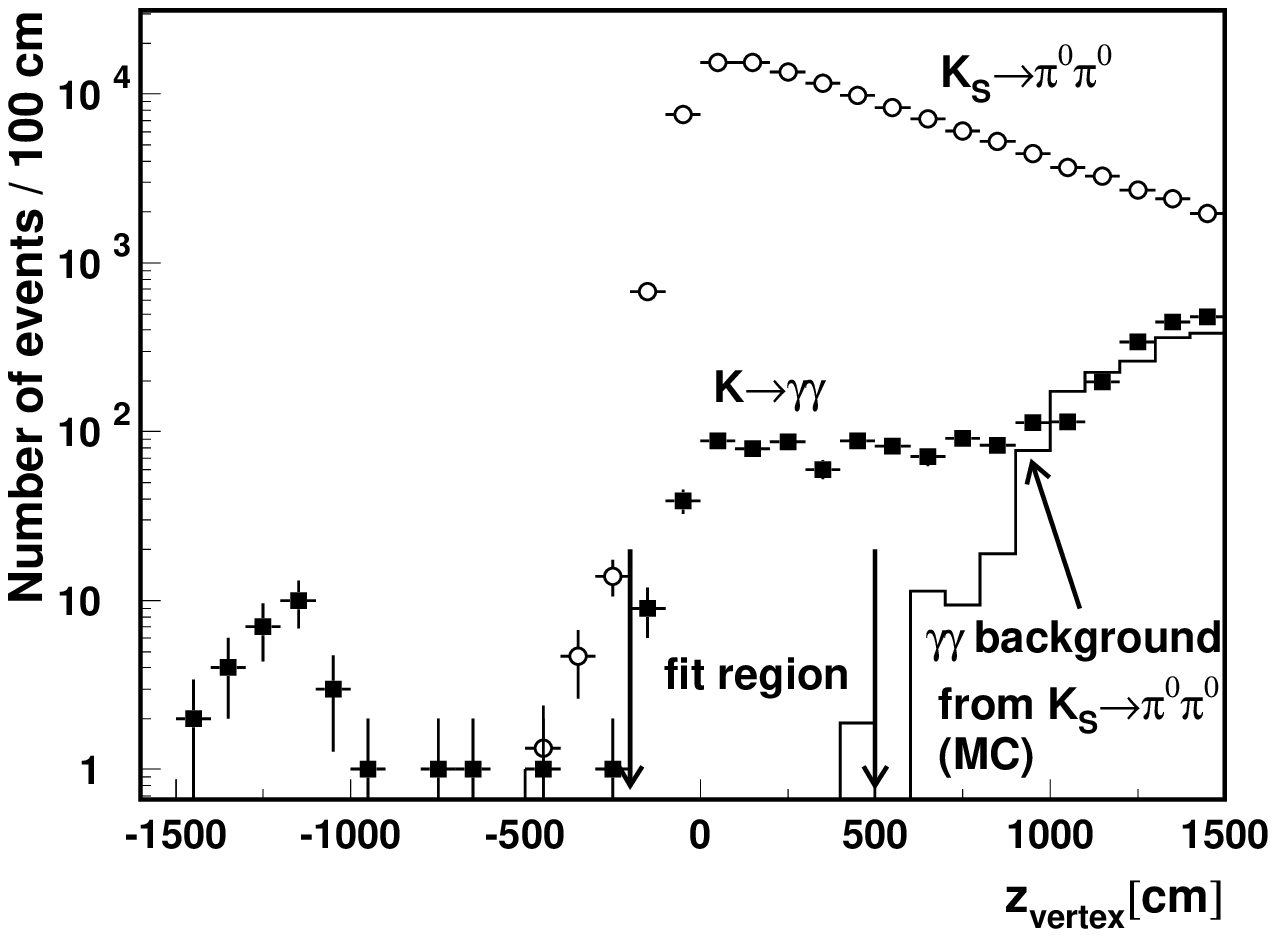,height=6cm}
\hskip 1.0cm
\psfig{figure=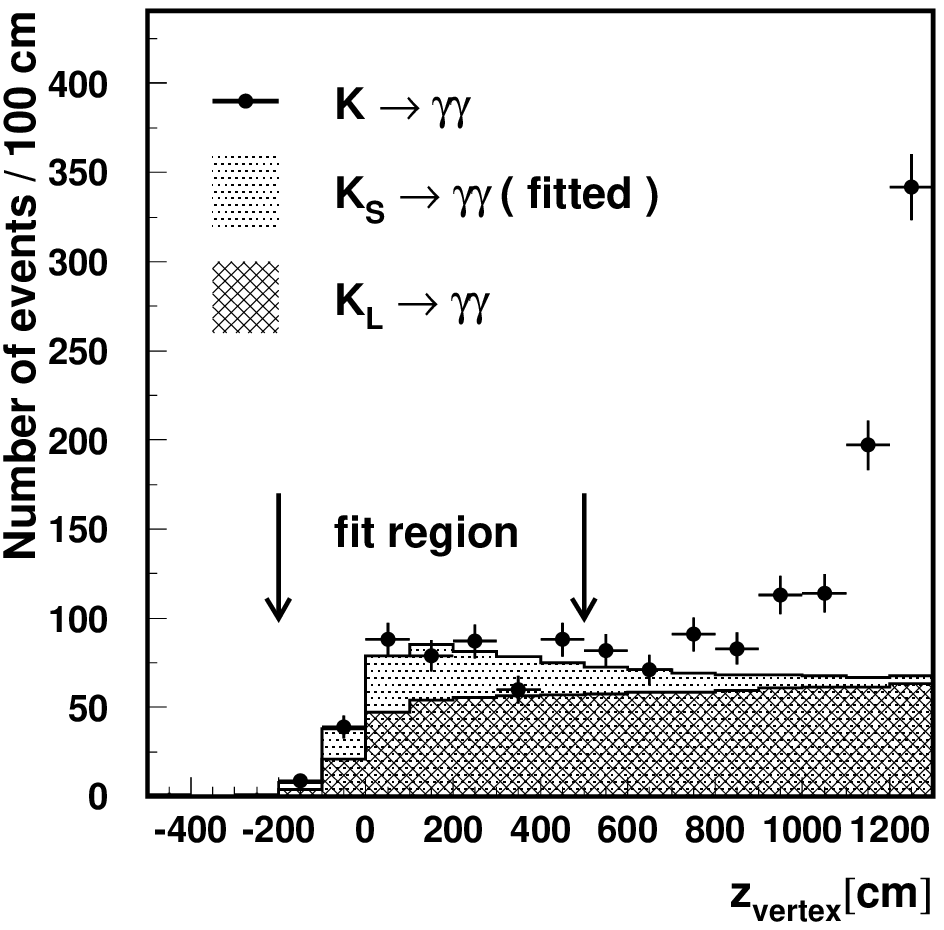,height=6cm}
\caption{\it (left) The $z_{vertex}$ distribution for data and MC. 
The open circles represent
$\kspp$ events while closed squares show $\kgg$. 
The solid line shows the $\gamma\gamma$
background from MC simulated $\kspp$ decays. The peak at $z=-1250$ cm is due
to $\eta$ mesons produced in the AKS. 
(right) The $z_{vertex}$ distribution obtained from maximum likelihood method
 for fitted $\ksgg$ events (dotted area) and $\klgg$ component (hatched area).
The dots show the data and the arrows show the fitted region. }
\label{fig:zkgg}
\end{figure}

Two-gamma decays of short-lived
kaons are reconstructed from $\ks$ high intensity data taken in 1999.
$\ksgg$ candidates are selected by requiring events with $\geq2$ clusters. The 
pairs of these clusters must be within $\pm5$ ns. No other cluster with energy 
$>1.5$ GeV within $\pm3$ns around the event time must be present. The energy of 
each cluster must lie between $3<E_{i}<100$ GeV and the sum of cluster energies, 
the kaon energy, must be between 60 and 170 GeV. The center of gravity(cog) measured 
for two clusters $cog = {\sqrt{(\sum_i E_i x_i)^2+(\sum_i E_i y_i)^2}}/{\sum_i 
E_i}$, must be less than 7cm. Events with some activity in the AKS, 
AKL and DCH are rejected. Energy deposited in the HAC must 
not exceed 3 GeV in a time window of $\pm15$ns around the event time. 
After the initial event selection, the longitudinal vertex position of the 
event is calculated by 
\beq z_{vertex}= z_{LKR} -
\frac{\sqrt{\sum_{i,j,i>j}{E_i}{E_j} [(x_i - x_j)^2 + (y_i - y_j)^2]}}{m_K}, 
\nonumber \eeq
where $z_{LKR}$ is the longitudinal position of the liquid krypton calorimeter
from the AKS, $x_i, y_i$ are the transverse position of the clusters at the LKR, and
$m_K$ is the kaon mass. 
The study of $\ksgg$ is done in a limited region, 
$-2<z_{vertex}<5$ m in order to minimize the main source of
background which is $\kspp$. After the above cuts, 450 $\kgg$ events are
obtained. The $z_{vertex}$ distribution of those events are shown in Figure~\ref{fig:zkgg}(left).

The main background to $\ksgg$ decays are $\kspp$ decays where
two photons are not detected.
The maximum invariant mass of the $\gamma\gamma$
from $\pi^0\pi^0$ events is 458 MeV resulting 
in a 9 m shift in the $z_{vertex}$. This background is 
minimized by requiring the $z_{vertex}$ cut described above.
$10^8$  $\kspp$ events are simulated to estimate this contribution. 
A second background to $\ksgg$ decays arises from $\klgg$ decays because of 
the $\kl$ flux produced at the target. This background is
estimated by measuring $\kl$ flux from $\kspp$ decays.
The remaining background originated from $\lnp$ is estimated by
comparing cog distribution for those events with signal events. 

Using a binned maximum likelihood method, $149\pm21$ $\ksgg$ events 
are estimated in the signal region and the branching ratio 
of $\ksgg$ is measured to be~\cite{gg_na48}
\beq
BR(K_{S}\rightarrow\gamma\gamma) = (2.58 \pm0.36(stat) \pm0.22(sys))\times10^{-
6}.
\eeq
The result of the likelihood method is shown in Figure~\ref{fig:zkgg}(right).
The main systematic 
error sources are due to uncertainty of $BR(\klgg)(5\%)$, 
the selection cuts ($4\%$), background ($5\%$), 
acceptance ($2\%$), and the trigger efficiency ($2\%$).

From this measurement, the ratio of the decay widths of $\ksgg$ to 
$\klgg$ is computed to be
\begin{center}
$R = \frac{\Gamma(\ksgg)}{\Gamma(\klgg)} = 2.53\pm0.35(stat)\pm0.22(sys).$
\end{center}

\section{Decays of $\klppee$}
The study of $\klppee$ decays~\cite{klppeeth,ksppeeth} 
provides a way to observe
CP violation in the neutral kaon system. The two main contributions to this
decay are the CP conserving direct emission associated
with a magnetic dipole transition (M1)
and the CP violating innner bremsstrahlung.
The interference between these two components
produces CP violating asymmetry ($A_{L}\sim14\%$) in the angle $\phi$ between normals
to the $\pi^{+}\pi^{-}$ and $e^{+}e^{-}$ planes 
in the kaon center of mass system.

The asymmetry and the branching ratio is determined from the combined
data taken in 1998 and 1999. $\klppee$ events are selected by requiring 
two positive and two negative intime tracks. The electron and pion
identifications are based on the $E/p$ measurement,
where $E$ is measured in the LKR and $p$ is measured in the spectrometer. 
 
\begin{figure}[ht]
\psfig{figure=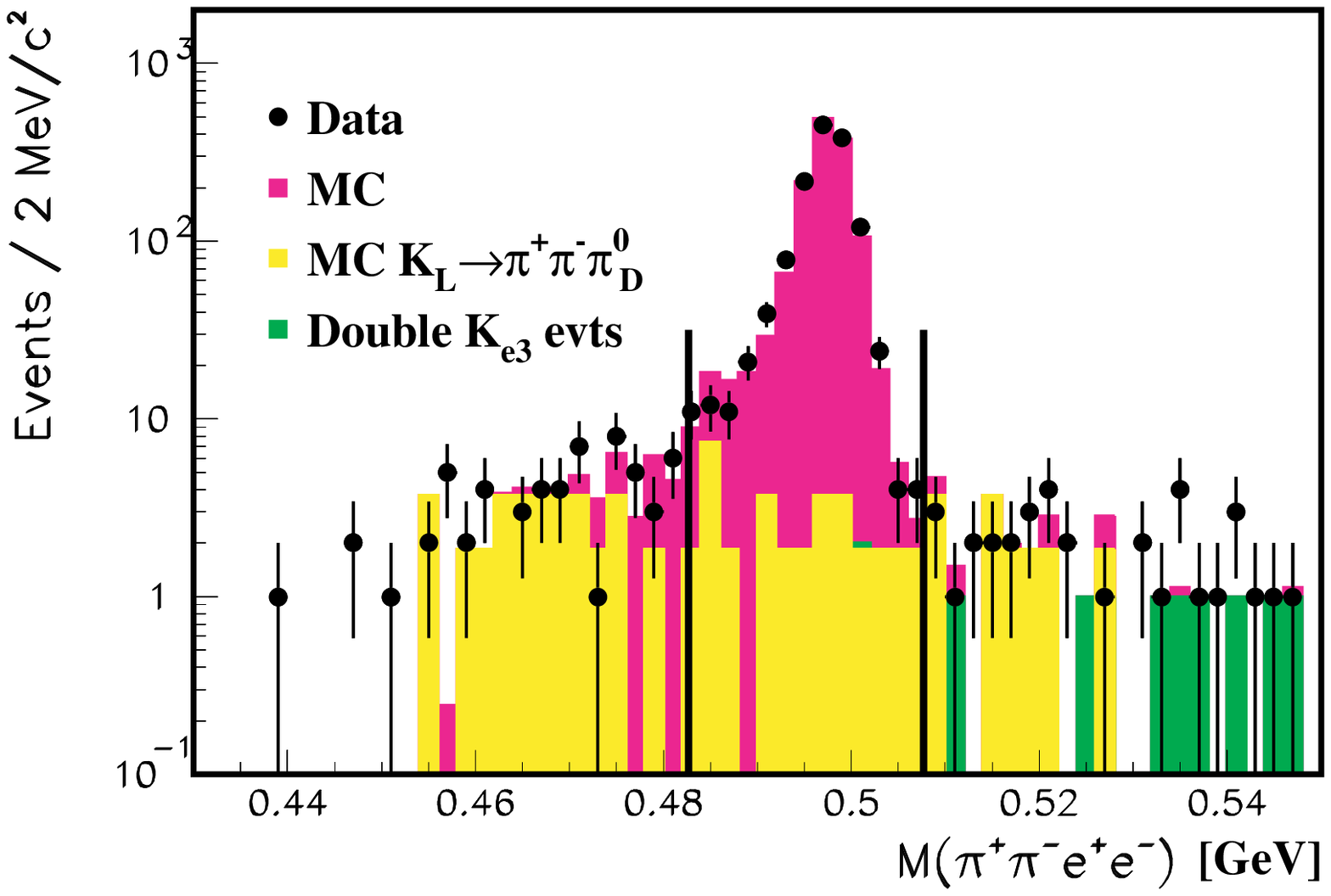,height=5cm}
\hskip 1.0cm
\psfig{figure=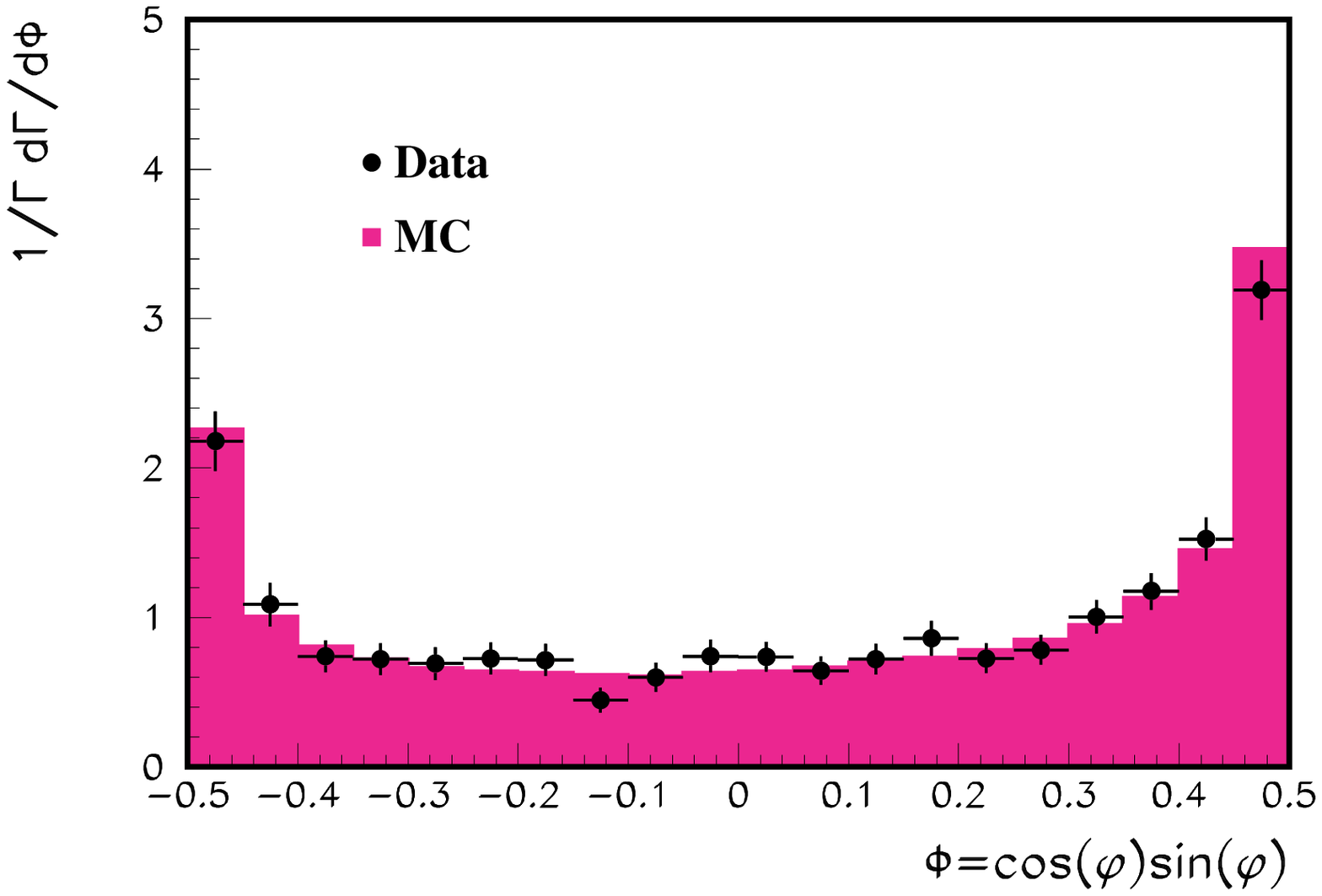,height=5cm}
\caption{\it For combined 1998 and 1999 data: (left) Measured $M_{\pi^{+}\pi^{-}e^{+}e^{-}}$ invariant mass for $\klppee$, 
(right) After acceptance correction, the $\phi$ distribution from which asymmetry is obtained.}
\label{fig:mklppee}
\end{figure}

The main background to $\klppee$ decays is due to the $\klpppd$ where
$\pi^{0}_{D}$ decays into $e^{+}e^{-}\gamma$, with a lost $\gamma$.
They are eliminated using a variable $p^{'2}_{0}$,
$p^{'2}_{0} = ((m_{K}^{2}-m_{\pi^{0}}^2-m_{\pi\pi}^2)^{2}-4m_{\pi^{0}}^2m_{\pi\pi}^2-
4(p_{\perp}^{2})_{\pi\pi}m_{K}^{2}) / 
4(m_{\pi\pi}^2+(p_{\perp}^{2})_{\pi\pi})$.  
For $\klpppd$ events $p^{'2}_{0}$ is greater than zero.
The second background source is due to the $\klppg$, where $\gamma$ 
converts into a $e^{+}e^{-}$ pair. The rejection is done by 
demanding 2 cm seperation between electron
tracks in the first drift chamber.

The invariant mass $m_{\pi^{+}\pi^{-}e^{+}e^{-}}$ distribution 
for $\klppee$ is shown in Figure~\ref{fig:mklppee}(left).
Using KTEV results~\cite{klppee_ktev} on $a_{1}/a_{2}=-0.72\pm0.03$ and
$\tilde{g}_{M1}=1.35^{+0.20}_{-0.17}$, 
the preliminary measurement on branching ratio of $\klppee$, based on 1337
reconstructed events, is 
$BR(K_{L}\rightarrow\pi^{+}\pi^{-}e^{+}e^{-}) = (3.1 \pm0.1(stat) \pm0.2(sys))\times10^{-7}.$
The asymmetry is obtained from the $\phi$ distribution corrected 
for acceptance (Figure~\ref{fig:mklppee}(right)). The preliminary result on the asymmetry
is given by
$A_{L} = (13.9 \pm2.7(stat) \pm2.0(sys))\%$
which is in good agreement with the theoretical prediction.

\section{Decays of $\ksppee$}

The main contribution to the $\ksppee$ decay~\cite{klppeeth,ksppeeth}
arises from CP conserving inner bremsstrahlung and as a result
no asymmetry in the $\phi$ distribution
is expected. 

The events selection is done in a similar way as in $\klppee$. 
The main background arising from $\klpppd$ is eliminated
by requiring the event to be tagged as coming from the $\ks$ beam.
The first observation of this decay is based on the data taken in 
1998~\cite{ksppee}. Based on the 56 $\ksppee$ events,  
the value obtained for the branching ratio of $\ksppee$ is
$BR(K_{S}\rightarrow\pi^{+}\pi^{-}e^{+}e^{-}) = (4.5 \pm0.7(stat) \pm0.4(sys))\times10^{-5}.$
This value translates into the inner bremsstrahlung component
of $\klppee$, $BR(K_{L}^{IB}\rightarrow\pi^{+}\pi^{-}e^{+}e^{-}) =
(1.4 \pm0.2)\times10^{-7}$.

\begin{figure}[ht]
\psfig{figure=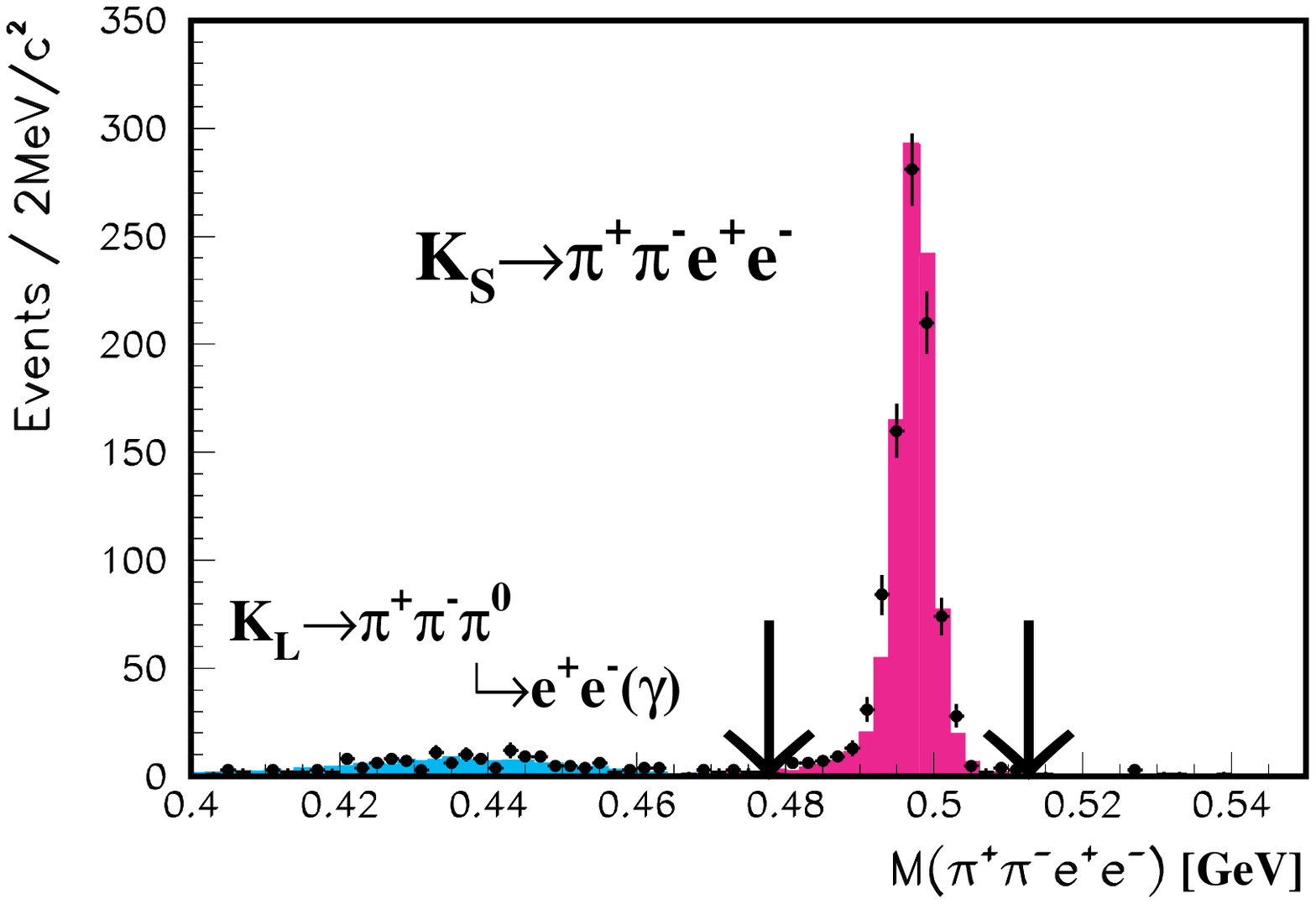,height=5cm}
\hskip 1.0cm
\psfig{figure=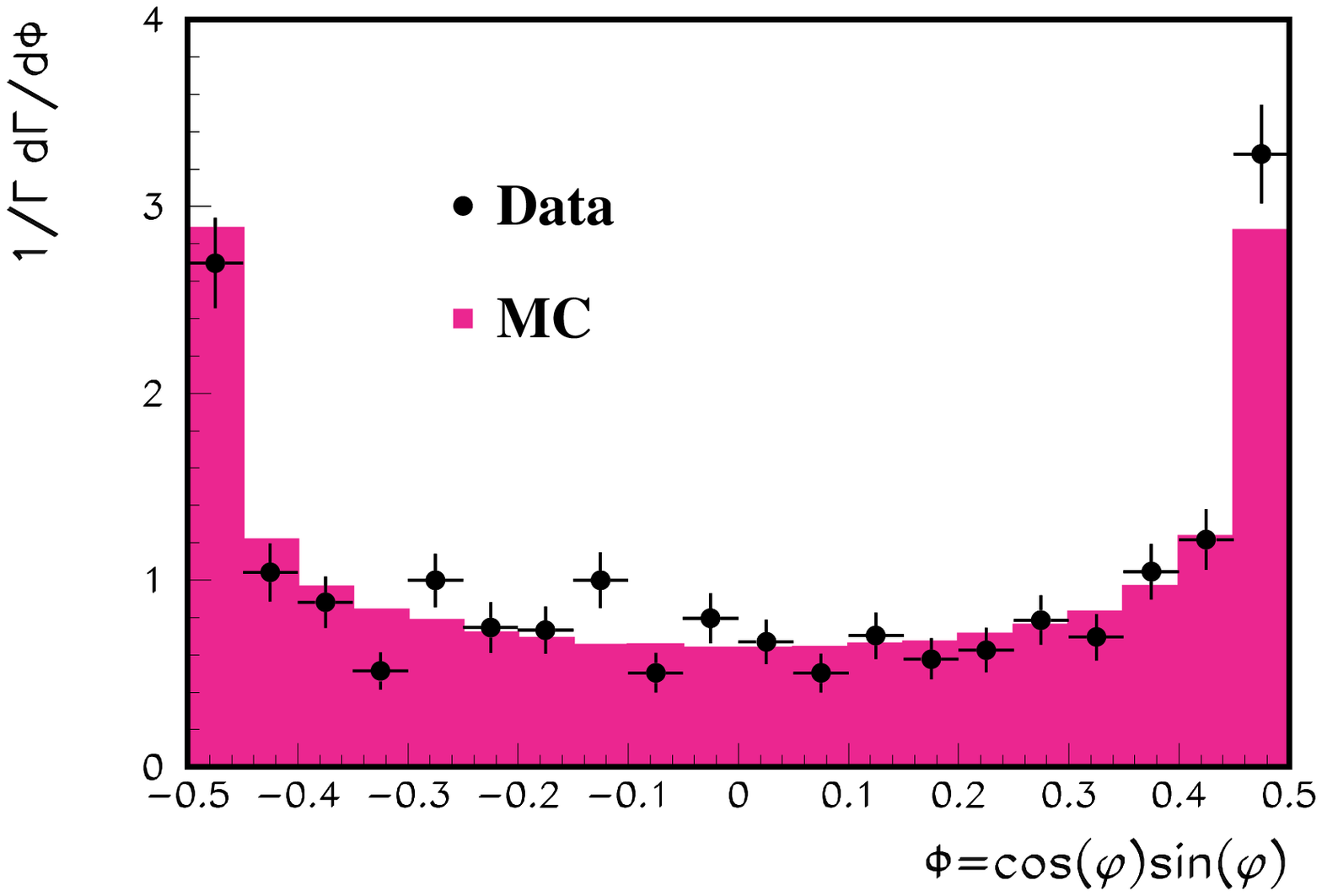,height=5cm}
\caption{\it For combined 1998 and 1999 data: (left) Measured $M_{\pi^{+}\pi^{-}e^{+}e^{-}}$ invariant mass for $\ksppee$, 
(right) After acceptance correction, the $\phi$ distribution from which asymmetry is obtained.}
\label{fig:mksppee}
\end{figure}

The asymmetry, $A_{S}$, and branching ratio of $\ksppee$ are measured
by using combined 1998 and 1999 data including the 2-day $\ks$
high intensity data.
Figure~\ref{fig:mksppee} shows the invariant mass $M_{\pi^{+}\pi^{-}e^{+}e^{-}}$ 
and $\phi$ distributions for these data.
Based on 921 $\ksppee$ events, the preliminary result for the 
asymetry is given by
$A_{S} = (-0.2\pm3.4(stat)\pm1.4(sys))\%$ which is consistent with zero. 
The value obtained for the branching ratio  is 
$BR(K_{S}\rightarrow\pi^{+}\pi^{-}e^{+}e^{-}) = (4.3 \pm0.2(stat)\pm0.3(sys))
\times10^{-5}$, and the inner bremsstrahlung component
of $\klppee$,  $BR(K_{L}^{IB}\rightarrow\pi^{+}\pi^{-}e^{+}e^{-}) = 
(1.3 \pm0.1))\times10^{-7}$.

\section{Decays of $\kspee$}

The study of $\kspee$ decay~\cite{kspee_th} is important to improve the limit on the 
indirect CP violating term in $\klpee$. The branching ratio is given by:
$BR(K_{S}\rightarrow\pi^{0}e^{+}e^{-}) \sim 5.2\times10^{-9}a_{S}^{2}$
where $a_{S}$ is the strength of the indirect CP violating 
component in $\klpee$.  The present limit on the branching ratio comes from
the NA31 experiment~\cite{kspee_na31}, $BR(K_{S}\rightarrow\pi^{0}e^{+}e^{-}) < 1.1\times10^{-6}$
at 90\% confidence level. A more precise measurement of branching ratio of $\kspee$
will provide better limits on the indirect CP violating contribution in $\klpee$.

The $\kspee$ search is based on the $\ks$ high intensity data taken in 1999.
The events are selected by requiring at least four cluster in the LKR, 
two tracks and one vertex. Then the group of four clusters has to pass
additional kinematic cuts such as the invariant mass of $\gamma\gamma$ must be
within 2.5 MeV with respect to the nominal $\pi^0$ mass, and the invariant mass
of $ee\gamma\gamma$ must lie within 10 MeV with respect to nominal kaon mass. 

The main background $\kspee$ comes from $\kspdpd$, where both $\pi^{0}_{D}$
decays into $e^{+}e^{-}\gamma$, and one electron and one positron are lost. This background
is rejected by requiring invariant mass difference $|m_{e\gamma}-m_{\pi^0}|$ to be greater than 30 MeV.
Another source of background arises from $\ksppd$, where $\pi^{0}_{D}$
decays into $e^{+}e^{-}\gamma$ with a lost $\gamma$. Due to the origin of the
electrons, their invariant mass cannot exceed pion mass. The remaining background in $\kspee$
are negligible.

\begin{figure}[ht]
\begin{center}
\psfig{figure=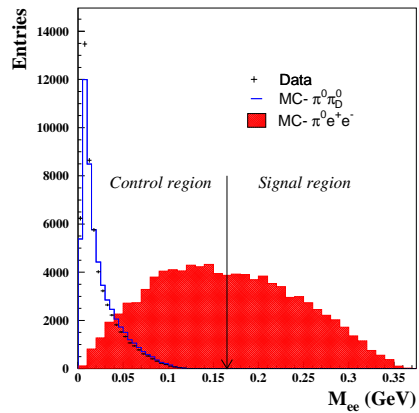,height=6cm}\end{center}
\caption{\it Invariant mass distribution of $e^{+}e^{-}$ for data (dots) and
for $\pi^{0}\pi^{0}_{D}$ monte-carlo events (solid line). The expected distribution
for $\kspee$ events is shown by the shaded area.}
\label{fig:mkspee}
\end{figure}

Figure~\ref{fig:mkspee} shows the invariant mass
distribution of $e^{+}e^{-}$ pair for events passing the above cuts.
The events below 
$m_{ee} < 165 $ MeV  is in good agreement with the background
$\ksppd$ events. No events are observed $m_{ee}$ above 165 MeV. 
Using $\ksppd$ as a 
normalization channel, at 90\% confidence level upper limit on the
branching ratio is obtained, 
\beq
BR(K_{S}\rightarrow\pi^{0}e^{+}e^{-}) < 1.4\times10^{-7}.
\eeq 

\end{document}